\documentstyle[12pt,fleqn]{article}
\input psfig

\catcode`\@=11
\@addtoreset{equation}{section}

\def\be{\begin{equation}}
\def\ee{\end{equation}}
\def\ba{\begin{eqnarray}}
\def\ea{\end{eqnarray}}

\def\fun#1#2{\lower3.6pt\vbox{\baselineskip0pt\lineskip.9pt
        \ialign{$\mathsurround=0pt#1\hfill##\hfil$\crcr#2\crcr\sim\crcr}}}

\def\Pbf{\mbox{\boldmath$P$}}
\def\Sbf{\mbox{\boldmath$S$}}
\def\Qbf{\mbox{\boldmath$Q$}}

\begin{document}
\begin{titlepage}
\null\vspace{-70pt}
\begin{flushright}FERMILAB-Pub-95/383-A\\
astro-ph/9512140\\
Submitted to {\it Physical Review D}\\
December 1995
\end{flushright}

\vspace{.2in}
\baselineskip 24pt
\centerline{\Large \bf A FAST AND ACCURATE ALGORITHM}
\centerline{\Large \bf FOR COMPUTING TENSOR CBR ANISOTROPY}

\vspace{.3in}
\centerline{Michael S. Turner$^{1,2}$ and Yun Wang$^1$}
\vspace{.2in}
\centerline{{\it $^1$NASA/Fermilab Astrophysics Center}}
\centerline{\it Fermi National Accelerator Laboratory, Batavia, IL~~60510-0500}
\vspace{.1in}
\centerline{\it Departments of Physics and of Astronomy \& Astrophysics,
Enrico Fermi Institute}
\centerline{\it The University of Chicago, Chicago, IL~~60637-1433}

\vspace{.4in}
\centerline{\bf Abstract}
\begin{quotation}

Inflation gives rise to a nearly scale-invariant spectrum of
tensor perturbations (gravitational waves), their contribution
to the Cosmic Background Radiation (CBR) anisotropy depends upon
the present cosmological parameters as well as inflationary parameters.
The analysis of a sampling-variance-limited CBR map offers the most
promising means of detecting tensor perturbations, but will require
evaluation of the predicted multipole spectrum for a very large
number of cosmological parameter sets.
We present accurate polynomial formulae for computing the predicted
variance of the multipole moments in terms of the cosmological
parameters $\Omega_\Lambda$, $\Omega_0h^2$, $\Omega_B h^2$, $N_{\nu}$,
and the power-law index $n_T$ which are accurate to about 1\% for
$l\le 50$ and to better than 3\% for $50< l \le 100$ (as compared
to the numerical results of a Boltzmann code).

\end{quotation}

PACS index numbers: 04.30.+x, 98.80.Cq, 98.80.Es

\end{titlepage}

\baselineskip=24pt

\section{I. Introduction}

Inflation makes three robust predictions:  flat Universe and
nearly scale-invariant spectra of scalar (density)
and tensor (gravitational-wave) metric perturbations \cite{inflate}.
The scalar \cite{scalar} and tensor \cite{tensor} perturbations
arise from quantum mechanical
fluctuations on very small scales during inflation and
are stretched to astrophysically interesting scales by the
tremendous growth of the cosmic scale factor during inflation.

Both the scalar and tensor perturbations give rise to anisotropy
in the temperature of the Cosmic Background Radiation (CBR) seen
on the sky today, most conveniently described by their contribution to the
multipole decomposition of the CBR temperature
\begin{equation}
\delta T(\theta ,\phi ) /T = \sum_{l=2}^{\infty} \sum_{m=-l}^{l}
        a_{lm} Y_{lm}(\theta ,\phi ) .
\end{equation}
The scalar and tensor contributions to the
anisotropy predicted by inflation
are uncorrelated and statistical in character.
The individual multipoles that describe our sky
are given by the sum of a scalar
plus tensor contribution, with these contributions
being drawn from gaussian distributions with variances
$\langle |a_{lm}^S|^2\rangle$ and $\langle |a_{lm}^T|^2\rangle$,
which are related to the properties of the inflationary potential
and cosmological parameters.  Because the scalar and tensor contributions
are uncorrelated, the variance $\langle |a_{lm}|^2\rangle
=\langle |a_{lm}^S|^2\rangle + \langle |a_{lm}^T|^2 \rangle$.
The expected scalar \cite{sugiyama} and
tensor contributions are shown in Fig.~1 for a nominal set
of cosmological parameters.

CBR anisotropy offers a very promising means of testing inflation
as well as determining the scalar and tensor perturbations.  If
both can be measured, then information about the underlying
inflationary potential can be derived (value of the potential
and its first few derivatives at a point) \cite{reconstruct}.
Key to doing this is a high-angular-resolution (better than $0.5^\circ$),
sampling-variance-limited map of the CBR sky.
(Because there are only $2l+1$
multipoles, sampling variance limits the accuracy to which
$\langle |a_{lm}|^2 \rangle$ can be measured -- a relative
precision of $\sqrt{2/(2l+1)}$.)  Three proposals have been
made to NASA (FIRE, PSI, and MAP) for a satellite-borne experiment
and another to ESA (COBRAS/SAMBA). A satellite could
be launched as early as 1999.

The separation of the scalar and tensor contributions
to CBR anisotropy is likely to be done by maximum likelihood techniques
and will require accurate predictions for the scalar and tensor contributions
(1\% or better) to the anisotropy for many sets of cosmological
and inflationary parameters. At present, achieving such precision
requires numerically integrating Boltzmann equations,
which is very time consuming (typically requiring many
hours on a powerful workstation for one set of cosmological
parameters) \cite{deconstruct}. The need for a fast and accurate
approximation scheme is manifest.

Many analytic approximations to the tensor angular power spectrum
have been explored \cite{tensorcalc}.  The most accurate
is as time consuming as numerically integrating the Boltzmann
equations \cite{allenkoranda}, and even schemes with less
accuracy require significant computation time (tens of minutes).  This
motivated the present approach -- a polynomial fit that can
be evaluated very rapidly (much less than a second for a set
of cosmological parameters).

The tensor angular power spectrum, $C_l(\Pbf)
= \langle |a_{lm}^T|^2\rangle$,
depends upon a set of cosmological and inflationary parameters
denoted here by $\Pbf$.  The set $\Pbf$ is:  the baryon density,
$\Omega_Bh^2$; the matter density, $\Omega_0h^2$; the level
of radiation in the Universe, parameterized by the equivalent number of
massless neutrino species $N_\nu$; the present vacuum-energy density,
$\Omega_\Lambda\equiv 1-\Omega_0$; and the primordial power-law index of
the tensor perturbations, $n_T$ ($n_T=0$ for scale invariant tensor
perturbations).  We will only be concerned with the shape of the angular
power spectrum; the overall amplitude of the angular power spectrum,
conveniently specified by $C_2$, depends upon the inflationary parameters
(the value of the inflationary potential in Planck units) as
described elsewhere \cite{turnerwhite}. In most approaches,
the shape $(C_l/C_2)$ and the overall amplitude $(C_2)$ are
determined independently.

The dependence upon the parameters is simple to explain.
The shape of the angular power spectrum depends on the redshift of last
scattering and the evolution of the gravitational waves
after they enter the horizon, which depends upon the evolution of
the cosmic scale factor. The redshift of last scattering (more precisely,
the peak of the visibility function; see Appendix B) depends upon the baryon
density, the matter density, and the level of radiation in the Universe.
The evolution of the scale factor of the Universe
around last scattering depends upon the relative levels of
matter and radiation through the value of the scale
factor at matter-radiation equality,
\be
R_{\rm EQ} =4.16\times 10^{-5}(\Omega_0 h^2)^{-1}
\left(\frac{2+0.4542 N_\nu}{3.3626}\right).
\label{eq:Req}
\ee
The recent evolution of the scale factor,
which depends upon $\Omega_\Lambda$ as well, is also important for
gravitational wave modes which have recently entered the horizon and influence
the low $l$ multipoles.  Finally, the primordial spectral shape of tensor
perturbations (described by $n_T$) also affects the shape of the angular
power spectrum.

We have engineered our fits based upon the numerical results of
the Boltzmann code written by Dodelson and Knox \cite{tensorcode};
they believe that their code is accurate to better than 1\%.
We expand $C_l (\Pbf)$ around a fiducial set
of parameters:  for $\Omega_\Lambda = 0$ ($\Omega_0 =1$),
$h=0.5$, $N_\nu =3$, $\Omega_Bh^2 = 0.0125$, and $n_T=0$;
and for $\Omega_\Lambda \not= 0$, $\Omega_0h^2 = 0.125$,
$\Omega_B h^2 = 0.0125$, $\Omega_\Lambda = 0.5$, $N_\nu =3$
and $n_T=0$.  These two cases are treated in the next two
Sections.  We end with a brief discussion of the
accuracy of our fits.

\section{II. Zero Cosmological Constant}

For $\Omega_\Lambda =0$, the cosmological parameters
are $(h, N_{\nu}, \Omega_B h^2, n_T)$.
We define a parameter vector,
$\Pbf=\left([h\, f(N_{\nu})], N_{\nu}, [\Omega_B h^2]^{-1}, n_T\right)$,
where
\be
f(N_{\nu}) = \sqrt{\frac{ 3.3626}{2+0.4542\, N_{\nu}} },
\label{eq:fnu}
\ee
which is related to $R_{\rm EQ}$ of Eq.(\ref{eq:Req}).

We expand the tensor multipole spectrum for $\Pbf$
around the tensor multipole spectrum for
$\Pbf=\Qbf\equiv (0.5, 3, 80, 0)$ [which corresponds to
$h=0.5, N_{\nu}=3, \Omega_B h^2=0.0125$, and $n_T=0$]. We write
\be
\frac{C_l(\Pbf)}{C_2(\Pbf)}=
\frac{C_l(\Qbf)}{C_2(\Qbf)}\,\left(\frac{l}{2}\right)^{n_T}
\left[1+g(l)\, n_T\right]\,\left\{ \frac{1+\sum_{i=1}^3 f_i(P_i-Q_i, l)}
{1+\sum_{i=1}^3 f_i(P_i-Q_i, 2)}\right\}.
\ee
where
\ba
&& f_1(P_1-Q_1, l) = \sum_{j=1}^4 a_j(l)\, (P_1-Q_1)^j \nonumber\\
&& f_2(P_2-Q_2, l) = \sum_{j=1}^3 b_j(l)\, (P_2-Q_2)^j \nonumber\\
&& f_3(P_3-Q_3, l) = \sum_{j=1}^3 d_j(l)\, (P_3-Q_3)^j
\ea
The coefficients are found numerically by fitting to the variation
due to each parameter separately. Identifying the relevant
vector of parameters minimizes the need for cross terms.

The $P_1$ coefficients are
\ba
a_1(l)&=&  -0.4025 \left(\frac{l}{100}\right)
 -0.3375 \left(\frac{l}{100}\right)^2
  -2.3441	\left(\frac{l}{100}\right)^3
  +2.0125	\left(\frac{l}{100}\right)^4 \nonumber \\
a_2(l)&=& 1.1271 \left(\frac{l}{100}\right)
  -2.0614 \left(\frac{l}{100}\right)^2
  +10.9105	\left(\frac{l}{100}\right)^3
 -10.8102	\left(\frac{l}{100}\right)^4 \nonumber \\
a_3(l)&=&   -2.7875 \left(\frac{l}{100}\right)
+10.9417 \left(\frac{l}{100}\right)^2
  -36.4181	\left(\frac{l}{100}\right)^3
 +41.1617	\left(\frac{l}{100}\right)^4 \nonumber\\
a_4(l)&=&   3.4719 \left(\frac{l}{100}\right)
-16.8888 \left(\frac{l}{100}\right)^2
  +51.2793	\left(\frac{l}{100}\right)^3
 -61.1775	\left(\frac{l}{100}\right)^4. \nonumber\\
&&
\ea
The $P_2$ coefficients are
\ba
b_1(l)&=&  10^{-3} \left[
-0.0361 \left(\frac{l}{100}\right)+5.3684  \left(\frac{l}{100}\right)^2
 +3.3680 \left(\frac{l}{100}\right)^3 \nonumber \right.\\ &&\hskip 3cm\left.
-2.0872 \left(\frac{l}{100}\right)^4 +1.2639 \left(\frac{l}{100}\right)^5
\right]\nonumber \\
b_2(l)&=& 10^{-3} \left[
-0.1105 \left(\frac{l}{100}\right) +0.1237 \left(\frac{l}{100}\right)^2
  -1.8853 \left(\frac{l}{100}\right)^3 \nonumber \right.\\ &&\hskip 3cm\left.
 +2.2411 \left(\frac{l}{100}\right)^4-1.0243\left(\frac{l}{100}\right)^5
\right] \nonumber \\
b_3(l)&=& 10^{-4} \left[
0.1297 \left(\frac{l}{100}\right)-0.5125 \left(\frac{l}{100}\right)^2
+2.3069\left(\frac{l}{100}\right)^3 \nonumber \right.\\ &&\hskip 3cm\left.
-2.9658	\left(\frac{l}{100}\right)^4 +1.5493 \left(\frac{l}{100}\right)^5
\right].\nonumber\\
&&
\ea

The $P_3$ coefficients are
\ba
d_1(l)&=&  10^{-4} \left[
 -0.1972 \left(\frac{l}{100}\right)
 + 12.3357  \left(\frac{l}{100}\right)^2
   -0.7469 	\left(\frac{l}{100}\right)^3
 +4.0362 \left(\frac{l}{100}\right)^4 \right]\nonumber \\
d_2(l)&=& 10^{-6} \left[
  -0.0241 \left(\frac{l}{100}\right)
  -1.7123 \left(\frac{l}{100}\right)^2
  -1.1385	\left(\frac{l}{100}\right)^3
 +0.6791	\left(\frac{l}{100}\right)^4\right] \nonumber \\
d_3(l)&=& 10^{-10} \left[
0.2028 \left(\frac{l}{100}\right)
+12.3878 \left(\frac{l}{100}\right)^2
   +8.9878	\left(\frac{l}{100}\right)^3
  -6.1211	\left(\frac{l}{100}\right)^4 \right].\nonumber\\
&&
\ea

The coefficient involving $n_T$ is
\ba
g(l)&=&   -0.4764\, \left[1- e^{-(l-2)/30}\right]
		+1.6734 \,\left[1- e^{-2(l-2)/30}\right]
	-4.0400	\,\left[1- e^{-3(l-2)/30}\right]\nonumber\\
&& \hskip 0.5cm +4.6345 \,\left[1- e^{-4(l-2)/30}\right]
	-2.2942 \,\left[1- e^{-5(l-2)/30}\right].
\ea

\section{III. Nonzero Cosmological Constant}

For $\Omega_\Lambda >0$, the cosmological parameters are ($\Omega_0 h^2$,
$\Omega_B h^2$, $\Omega_{\Lambda}$, $N_{\nu}$, $n_T$). The procedure
of fitting is similar to the $\Omega_{\Lambda}=0$
case, but slightly more complicated because one ``cross term'' is required
to achieve sufficient accuracy.

Here we define the parameter vector, $\Pbf=\left(\left[f(N_{\nu})
\sqrt{\Omega_0 h^2}\right], N_{\nu}, [\Omega_B h^2]^{-1}, \Omega_{\Lambda},
n_T\right)$. Again, we expand the $C_l$'s for $\Pbf$ around the $C_l$'s
for a standard set of cosmological parameters, $\Sbf$.
For $\Omega_{\Lambda}< 0.36$, we take $\Sbf=(0.5, 3, 80, 0, 0)$
[which corresponds to $h=0.5, N_{\nu}=3, \Omega_B h^2=0.0125,
\Omega_{\Lambda}=0$, and $n_T=0$].
For $0.36 \leq \Omega_{\Lambda}\leq 0.8$, we take $\Sbf=(0.3536, 3, 80,
0.5, 0)$ [which corresponds to $h=0.5, N_{\nu}=3, \Omega_B h^2=0.0125,
\Omega_{\Lambda}=0.5$, and $n_T=0$]. As before, we write
\ba
\frac{C_l(\Pbf)}{C_2(\Pbf)}=
\frac{C_l(\Sbf)}{C_2(\Sbf)}\,\left(\frac{l}{2}\right)^{n_T}
\left[1+\nu (l)\, n_T\right]\,\left\{ \frac{1+\sum_{i=1}^4 f_i(P_i-S_i, l)
+\eta (l) (P_1-S_1)(P_4-S_4)}
{1+\sum_{i=1}^4 f_i(P_i-S_i, 2)+\eta (2) (P_1-S_1)(P_4-S_4)}\right\},
\nonumber\\
\ea
where
\ba
f_1(P_1-S_1, l) &=& \sum_{j=1}^4 A_j(l)\, (P_1-S_1)^j \nonumber\\
f_2(P_2-S_2, l) &=& \sum_{j=1}^3 B_j(l)\, (P_2-S_2)^j \nonumber\\
f_3(P_3-S_3, l) &=&\sum_{j=1}^3 D_j(l)\, (P_3-S_3)^j \nonumber\\
f_4(P_4-S_4, l) &=&\sum_{j=1}^4 E_j(l)\, (P_4-S_4)^j,
\ea

For $\Omega_{\Lambda}< 0.36$,
\ba
&&A_j(l)=a_j(l), \hskip 0.5cm B_j(l)=b_j(l), \hskip 0.5cm
D_j(l)=d_j(l);
\nonumber\\
&&\nu(l)=g (l), \hskip 0.5cm \eta(l)=0.
\ea
The $P_4$ coefficients are
\ba
E_1(2)= -6.0163\times 10^{-2} ,\hskip 0.5cm &
E_1(3)= -1.4845\times 10^{-2} ,\hskip 0.5cm &
E_1(4)= -4.3594 \times 10^{-3} \nonumber \\
E_2(2)=  -3.2853\times 10^{-2},\hskip 0.5cm &
E_2(3)= -8.4852\times 10^{-3} ,\hskip 0.5cm &
E_2(4)= -2.5631\times 10^{-3} \nonumber \\
E_3(2)= -1.8426\times 10^{-2},\hskip 0.5cm &
E_3(3)= -4.0997\times 10^{-3} ,\hskip 0.5cm &
E_3(4)= -9.4267 \times 10^{-4} \nonumber \\
E_4(2)=  -2.8162\times 10^{-2},\hskip 0.5cm &
E_4(3)= -8.6348\times 10^{-3} ,\hskip 0.5cm &
E_4(4)=  -3.1561\times 10^{-3} \nonumber
\ea
\ba
E_1(l>4)&=& 0.1 \left[
0.1483 \left(\frac{l}{100}\right)
 -2.4098 \left(\frac{l}{100}\right)^2
  +0.3211\left(\frac{l}{100}\right)^3
 -0.6630\left(\frac{l}{100}\right)^4 \right] \nonumber \\
E_2(l>4)&=& 0.1 \left[
0.0768 \left(\frac{l}{100}\right)
 -1.5272 \left(\frac{l}{100}\right)^2
  -0.2895\left(\frac{l}{100}\right)^3
 +0.1754\left(\frac{l}{100}\right)^4 \right] \nonumber \\
E_3(l>4)&=& 0.1 \left[
0.0580\left(\frac{l}{100}\right)
 -1.0585 \left(\frac{l}{100}\right)^2
  -0.1897\left(\frac{l}{100}\right)^3
 +0.2662\left(\frac{l}{100}\right)^4 \right] \nonumber \\
E_4(l>4)&=& 0.1 \left[
0.0560 \left(\frac{l}{100}\right)
 -1.6921 \left(\frac{l}{100}\right)^2
  -0.8962\left(\frac{l}{100}\right)^3
 +1.0897\left(\frac{l}{100}\right)^4 \right]. \nonumber \\
&&
\ea
All other coefficients have been given in the previous Section.

For $0.36 \leq \Omega_{\Lambda}\leq 0.8$, the $P_1$ coefficients are
\ba
A_1(l)&=&  -1.0313\left(\frac{l}{100}\right)
+1.0270 \left(\frac{l}{100}\right)^2
 -7.6262 \left(\frac{l}{100}\right)^3
\nonumber\\&& \hskip 3cm
 +6.8176 \left(\frac{l}{100}\right)^4-2.9419\left(\frac{l}{100}\right)^5
\nonumber \\
A_2(l)&=& 3.5896 \left(\frac{l}{100}\right)
  -11.9704 \left(\frac{l}{100}\right)^2
 +37.1339	\left(\frac{l}{100}\right)^3 \nonumber\\&& \hskip 3cm
 -35.5609	\left(\frac{l}{100}\right)^4
+19.0353 \left(\frac{l}{100}\right)^5 \nonumber \\
A_3(l)&=& -7.6185 \left(\frac{l}{100}\right)
 +34.4566 \left(\frac{l}{100}\right)^2
  -92.1698	\left(\frac{l}{100}\right)^3 \nonumber\\&& \hskip 3cm
 +95.4331 \left(\frac{l}{100}\right)^4
-53.5697\left(\frac{l}{100}\right)^5 \nonumber\\
A_4(l)&=& 6.5272 \left(\frac{l}{100}\right)
 -32.9224 \left(\frac{l}{100}\right)^2
  +84.7298	\left(\frac{l}{100}\right)^3 \nonumber\\&& \hskip 3cm
 -91.4209 \left(\frac{l}{100}\right)^4
+51.8163\left(\frac{l}{100}\right)^5.
\ea

The $P_2$ coefficients are
\ba
B_1(l)&=&  10^{-2} \left[
-0.1629 \left(\frac{l}{100}\right)
 +1.4392  \left(\frac{l}{100}\right)^2
 -2.2268 \left(\frac{l}{100}\right)^3 \nonumber \right.\\ &&\hskip 3cm\left.
 +2.8450 \left(\frac{l}{100}\right)^4
-1.5396\left(\frac{l}{100}\right)^5 \right]\nonumber \\
B_2(l)&=&  10^{-4} \left[
0.3608\left(\frac{l}{100}\right)
 -6.3470  \left(\frac{l}{100}\right)^2
 +3.4559 \left(\frac{l}{100}\right)^3 \nonumber\right.\\
&& \left. \hskip 3cm
 -4.8620\left(\frac{l}{100}\right)^4
+7.0136 \left(\frac{l}{100}\right)^5 \right]\nonumber \\
B_3(l)&=&  10^{-4} \left[
0.1672 \left(\frac{l}{100}\right)
 -0.8800 \left(\frac{l}{100}\right)^2
 +2.9272\left(\frac{l}{100}\right)^3 \nonumber\right.\\ &&\left. \hskip 3cm
 -3.6758 \left(\frac{l}{100}\right)^4
+1.6453\left(\frac{l}{100}\right)^5 \right].
\ea

The $P_3$ coefficients are
\ba
D_1(l)&=&  10^{-4} \left[
-0.1039 \left(\frac{l}{100}\right)
 +9.2599  \left(\frac{l}{100}\right)^2
   +0.1085 \left(\frac{l}{100}\right)^3
 +2.4910 \left(\frac{l}{100}\right)^4 \right]\nonumber \\
D_2(l)&=& 10^{-7} \left[
-0.2716 \left(\frac{l}{100}\right)
   -16.4608 \left(\frac{l}{100}\right)^2
  -11.4085	\left(\frac{l}{100}\right)^3
+8.5966 \left(\frac{l}{100}\right)^4\right] \nonumber \\
D_3(l)&=& 10^{-9} \left[
0.0962 \left(\frac{l}{100}\right)
+2.8879 \left(\frac{l}{100}\right)^2
  +3.0975	\left(\frac{l}{100}\right)^3
-3.0516\left(\frac{l}{100}\right)^4 \right].\nonumber\\
&&
\ea

The $P_4$ coefficients are
\ba
E_1(2)= -1.3087\times 10^{-1} ,\hskip 0.5cm &
E_1(3)= -3.1368\times 10^{-2} ,\hskip 0.5cm &
E_1(4)=  -8.4146\times 10^{-3} \nonumber \\
E_2(2)= -1.1277 \times 10^{-1},\hskip 0.5cm &
E_2(3)= -2.7633\times 10^{-2} ,\hskip 0.5cm &
E_2(4)=  -7.3835\times 10^{-3} \nonumber \\
E_3(2)= -1.3674\times 10^{-1},\hskip 0.5cm &
E_3(3)= -3.4105\times 10^{-2} ,\hskip 0.5cm &
E_3(4)=  -8.9756\times 10^{-3} \nonumber \\
E_4(2)= -6.8693 \times 10^{-1},\hskip 0.5cm &
E_4(3)= -1.7588\times 10^{-1} ,\hskip 0.5cm &
E_4(4)=  -4.5389\times 10^{-2} \nonumber
\ea
\ba
E_1(l>4)&=& 0.1 \left[
0.5211 \left(\frac{l}{100}\right)
 -5.7831 \left(\frac{l}{100}\right)^2
  +1.0999\left(\frac{l}{100}\right)^3
 -1.2519\left(\frac{l}{100}\right)^4 \right] \nonumber \\
E_2(l>4)&=& 0.1 \left[
0.4547 \left(\frac{l}{100}\right)
 -5.2202 \left(\frac{l}{100}\right)^2
  -0.1399\left(\frac{l}{100}\right)^3
 -0.3364\left(\frac{l}{100}\right)^4 \right] \nonumber \\
E_3(l>4)&=& 0.1 \left[
0.5924\left(\frac{l}{100}\right)
 -7.3578 \left(\frac{l}{100}\right)^2
  -1.5359\left(\frac{l}{100}\right)^3
 +1.3647\left(\frac{l}{100}\right)^4 \right] \nonumber \\
E_4(l>4)&=&
0.5016 \left(\frac{l}{100}\right)
 -7.0197 \left(\frac{l}{100}\right)^2
  -0.2120\left(\frac{l}{100}\right)^3
 +2.9141\left(\frac{l}{100}\right)^4. \nonumber \\
&&
\ea

The coefficient involving $n_T$ is
\ba
\nu(l)&=&   -0.4670\, \left[1- e^{-(l-2)/30}\right]
		+ 1.7234 \,\left[1- e^{-2(l-2)/30}\right]
	-4.2724	\,\left[1- e^{-3(l-2)/30}\right]\nonumber\\
&& \hskip 0.5cm +4.9757 \,\left[1- e^{-4(l-2)/30}\right]
	-2.4571 \,\left[1- e^{-5(l-2)/30}\right].
\ea
Note that $\nu(l) \simeq g(l)$.

The cross-term coefficient is
\ba
\eta(l)&= &
-3.7201 \left(\frac{l}{100}\right)
+29.9770 \left(\frac{l}{100}\right)^2-117.6372 \left(\frac{l}{100}\right)^3
+236.1950 \left(\frac{l}{100}\right)^4 \nonumber\\
&& \hskip 1cm -257.9630 \left(\frac{l}{100}\right)^5+
147.3101 \left(\frac{l}{100}\right)^6
 -34.5437 \left(\frac{l}{100}\right)^7.
\ea

\section{IV. Discussion}

By identifying the relevant cosmological parameters
we have developed a fast ($\ll 1\,$sec) and accurate (few percent
or better) algorithm for computing the shape ($C_l/C_2$)
of the tensor angular power spectrum for a primordial tensor power
spectrum of the form
\begin{equation}
P_T(k) \ \propto \ (k\tau_0)^{n_T}k^{-3},
\end{equation}
where $\tau_0$ is the conformal time today.
Our algorithm employs a polynomial in the parameters
$\Omega_\Lambda$, $\Omega_0h^2$, $\Omega_Bh^2$, $n_T$, and $N_\nu$.

To assess the accuracy of our algorithm we sampled the following
parameter intervals uniformly and at random:
$0.35 \le h \le 0.8$, $2\le N_\nu \le 12$,
$0.005 \le \Omega_Bh^2 \le 0.03$, $-0.3 \le n_T \le 0$, and
$0 \le \Omega_\Lambda \le 0.8$.  In Figs. 2 and 3 we show the
histograms of the {\bf maximum error} in $C_l/C_2$ for $\Omega_\Lambda
=0$ and $\Omega_\Lambda \not= 0$ respectively.
For $\Omega_\Lambda =0$ the maximum error (relative to the Boltzmann code
of Ref.~\cite{tensorcode}) is less than about 0.5\% for
$l\le 50$ and less than about 2\% for $l\le 100$.  For $\Omega_\Lambda
\not= 0$, the accuracy is slightly worse, better than about
1\% for $l\le 50$ and better than about 3\% for $l\le 100$.
In the case of $\Omega_\Lambda \not= 0$ the largest errors
occur for $\Omega_0h^2 < 0.05$ (large $\Omega_\Lambda$ and
small $h$).

The tensor contribution to the quadrupole plays a special role.
It provides a convenient overall normalization for the angular
power spectrum and can be related to the value of the
inflationary potential when the comoving scale $k_* = H_0$
crossed outside the horizon during inflation \cite{turnerwhite}:
\begin{equation}
V_*/m_{\rm Pl}^4 = 0.66 \left[ 1. -(f_T^{(1)} +0.1)n_T\right]\,C_2/f_T^{(0)},
\end{equation}
where the functions $f_T^{(0,1)}(\Omega_\Lambda )$ are given by
\begin{eqnarray}
f_T^{(0)}(\Omega_\Lambda ) & = & 1. - 0.03\Omega_\Lambda
        -0.1\Omega_\Lambda^2  \nonumber \\
f_T^{(1)}(\Omega_\Lambda ) & = & 0.58 -0.50\Omega_\Lambda
        +0.31\Omega_\Lambda^2 -0.88\Omega_\Lambda^3.
\end{eqnarray}
The dependence of this relationship on cosmological
parameters other than $\Omega_\Lambda$ is much
less significant \cite{turnerwhite}.

There is even more motivation for developing a fast and accurate algorithm
for the scalar angular power spectrum.  However, this task is more challenging:
the power spectrum has more structure (cf., Fig. 1) and
that structure extends to higher multipoles.  We are currently working
on an algorithm for the scalar angular power spectrum.

\vskip 0.2in
\centerline{\bf Acknowledgments}
This work was supported by the DOE (at Chicago and at Fermilab)
and by the NASA through Grant NAG5-2788.

\section{Appendix A: Table of standard $C_l$'s}
\begin{tabular}{lll}
$l$ &  \large{$
\frac{l(l+1)C_l(\Omega_{\Lambda}=0)}{6C_2(\Omega_{\Lambda}=0)}$}  &
\large{$\frac{l(l+1)C_l(\Omega_{\Lambda}=0.5)}{6C_2(\Omega_{\Lambda}=0.5)}$ }\\
\hline\hline
 2&  1.000000000&  1.000000000 \\
 3&  7.85882861E-01&  8.12787313E-01 \\
 4&  7.43475686E-01&  7.75447410E-01 \\
 5&  7.35658928E-01&  7.69590303E-01 \\
 6&  7.38104089E-01&  7.73328636E-01 \\
 7&  7.43851511E-01&  7.80190352E-01 \\
 8&  7.50438650E-01&  7.87827791E-01 \\
 9&  7.56898572E-01&  7.95302374E-01 \\
 10&  7.62844076E-01&  8.02234328E-01 \\
 11&  7.68121692E-01&  8.08471560E-01 \\
 12&  7.72700749E-01&  8.13985567E-01 \\
 13&  7.76567875E-01&  8.18762161E-01 \\
 14&  7.79782672E-01&  8.22865837E-01 \\
 15&  7.82327767E-01&  8.26275290E-01 \\
 16&  7.84314505E-01&  8.29111207E-01 \\
 17&  7.85671940E-01&  8.31292902E-01 \\
 18&  7.86573958E-01&  8.33010933E-01 \\
 19&  7.86864814E-01&  8.34090814E-01 \\
 20&  7.86805032E-01&  8.34819817E-01 \\
 21&  7.86129624E-01&  8.34902088E-01 \\
 22&  7.85208901E-01&  8.34747737E-01 \\
 23&  7.83648115E-01&  8.33916120E-01 \\
 24&  7.81949251E-01&  8.32965076E-01 \\
 25&  7.79570004E-01&  8.31288581E-01 \\
 26&  7.77161094E-01&  8.29611689E-01 \\
 27&  7.74020055E-01&  8.27148151E-01 \\
 28&  7.70959238E-01&  8.24805533E-01 \\
 29&  7.67105366E-01&  8.21604444E-01 \\
 30&  7.63442668E-01&  8.18647318E-01 \\
 31&  7.58920711E-01&  8.14753604E-01 \\
 32&  7.54699914E-01&  8.11226404E-01 \\
 33&  7.49551094E-01&  8.06680893E-01 \\
 34&  7.44812364E-01&  8.02623603E-01 \\
 35&  7.39076276E-01&  7.97464933E-01 \\
 36&  7.33855970E-01&  7.92913085E-01 \\
 37&  7.27571031E-01&  7.87178514E-01 \\
 38&  7.21903458E-01&  7.82165288E-01 \\
 39&  7.15107322E-01&  7.75891111E-01 \\
 40&  7.09025177E-01&  7.70447667E-01 \\

\end{tabular}
\newpage
\begin{tabular}{lll}
$l$ & \large{
$\frac{l(l+1)C_l(\Omega_{\Lambda}=0)}{6 C_2(\Omega_{\Lambda}=0)}$}  &
\large{$\frac{l(l+1)C_l(\Omega_{\Lambda}=0.5)}{6C_2(\Omega_{\Lambda}=0.5)}$} \\
\hline\hline
 41&  7.01754819E-01&  7.63669098E-01 \\
 42&  6.95290317E-01&  7.57825769E-01 \\
 43&  6.87581696E-01&  7.50576759E-01 \\
 44&  6.80766517E-01&  7.44363318E-01 \\
 45&  6.72655788E-01&  7.36677916E-01 \\
 46&  6.65520307E-01&  7.30122320E-01 \\
 47&  6.57043916E-01&  7.22034754E-01 \\
 48&  6.49618128E-01&  7.15164612E-01 \\
 49&  6.40812380E-01&  7.06709069E-01 \\
 50&  6.33125395E-01&  6.99550707E-01 \\
 51&  6.24026867E-01&  6.90761689E-01 \\
 52&  6.16107568E-01&  6.83341454E-01 \\
 53&  6.06751841E-01&  6.74252272E-01 \\
 54&  5.98629552E-01&  6.66596862E-01 \\
 55&  5.89051404E-01&  6.57240302E-01 \\
 56&  5.80754839E-01&  6.49375955E-01 \\
 57&  5.70988919E-01&  6.39784406E-01 \\
 58&  5.62546297E-01&  6.31736935E-01 \\
 59&  5.52626894E-01&  6.21943006E-01 \\
 60&  5.44065284E-01&  6.13736759E-01 \\
 61&  5.34026679E-01&  6.03773187E-01 \\
 62&  5.25372061E-01&  5.95432062E-01 \\
 63&  5.15247817E-01&  5.85331082E-01 \\
 64&  5.06526007E-01&  5.76878240E-01 \\
 65&  4.96347985E-01&  5.66670612E-01 \\
 66&  4.87585435E-01&  5.58131161E-01 \\
 67&  4.77383343E-01&  5.47844948E-01 \\
 68&  4.68606069E-01&  5.39242990E-01 \\
 69&  4.58408817E-01&  5.28906521E-01 \\
 70&  4.49640932E-01&  5.20265325E-01 \\
 71&  4.39477440E-01&  5.09906642E-01 \\
 72&  4.30741650E-01&  5.01246850E-01 \\
 73&  4.20639523E-01&  4.90893728E-01 \\
 74&  4.11958163E-01&  4.82237697E-01 \\
 75&  4.01942685E-01&  4.71915135E-01 \\
 76&  3.93337874E-01&  4.63283465E-01 \\
 77&  3.83432932E-01&  4.53015076E-01 \\
 78&  3.74925150E-01&  4.44428914E-01 \\
 79&  3.65153926E-01&  4.34239289E-01 \\
 80&  3.56761877E-01&  4.25716630E-01 \\

\end{tabular}
\newpage
\begin{tabular}{lll}
$l$ &  \large{$
\frac{l(l+1)C_l(\Omega_{\Lambda}=0)}{6C_2(\Omega_{\Lambda}=0)}$ } &
\large{$\frac{l(l+1)C_l(\Omega_{\Lambda}=0.5)}{6C_2(\Omega_{\Lambda}=0.5)}$ }\\
\hline\hline
 81&  3.47146488E-01&  4.15627684E-01 \\
 82&  3.38887429E-01&  4.07185415E-01 \\
 83&  3.29449095E-01&  3.97220535E-01 \\
 84&  3.21338323E-01&  3.88874768E-01 \\
 85&  3.12097094E-01&  3.79055720E-01 \\
 86&  3.04149866E-01&  3.70821411E-01 \\
 87&  2.95121515E-01&  3.61164479E-01 \\
 88&  2.87354633E-01&  3.53060287E-01 \\
 89&  2.78552720E-01&  3.43581079E-01 \\
 90&  2.70980245E-01&  3.35623379E-01 \\
 91&  2.62418690E-01&  3.26338036E-01 \\
 92&  2.55052066E-01&  3.18539155E-01 \\
 93&  2.46743720E-01&  3.09462129E-01 \\
 94&  2.39593392E-01&  3.01833443E-01 \\
 95&  2.31549532E-01&  2.92978236E-01 \\
 96&  2.24624559E-01&  2.85530583E-01 \\
 97&  2.16855038E-01&  2.76910111E-01 \\
 98&  2.10163325E-01&  2.69653750E-01 \\
 99&  2.02676236E-01&  2.61279304E-01 \\
 100&  1.96225065E-01&  2.54224121E-01 \\

\end{tabular}
\section{Appendix B:  Last Scattering}

In the early Universe, matter and radiation were in good thermal contact,
because of the rapid interactions between the photons and electrons.
As the temperature dropped below 0.3eV, electrons
combined with protons to form neutral hydrogen (``recombination'')
at a redshift of around 1300.
With the disappearance of free electrons, the photon mean free path became
very large ($>H^{-1}$) and matter and radiation decoupled at a redshift of
around 1100. Last scattering is crucial in calculating the CBR anisotropy.
\cite{recombination}

The redshift of last scattering, $z_{\rm LSS}$, is given by the peak of the
visibility function $g(z)\equiv e^{-\tau} {\rm d}\tau/{\rm d z}$;
$g(z) {\rm d}z$ measures the probability that a given photon suffered
its last scattering in the redshift interval ($z$, $z+$d$z$).
The optical depth (measured from the present back to redshift $z$)
is given by
\be
\tau(z) = c \,\int^z_0 {\rm d}z \frac{{\rm d}t}{{\rm d}z}\, n_e(z)\,
 \sigma_T,
\ee
where ${\rm d}t $ is the proper time interval, $n_e(z)$ is the electron
number density, and $\sigma_T = 6.65\times 10^{-25}\,\mbox{cm}^2$ is
the Thomson scattering cross-section.
The electron number density depends on $\Omega_B h^2$ and $H(t)$,
the Hubble parameter at time $t$.
At the relevant times (around recombination and last scattering),
$H(t)$ only depends on $\Omega_0 h^2$ and $N_{\nu}$.
Hence, $z_{\rm LSS}$ only depends on
$\Omega_B h^2$, $\Omega_0 h^2$, and $N_{\nu}$.
Numerically, we find
\be
z_{\rm LSS}= 1104.37 + \Delta z(\Omega_B h^2,\Omega_0h^2)
+\Delta z(\Omega_0 h^2)
+\Delta z(N_{\nu},\Omega_B h^2, \Omega_0 h^2),
\ee
where
\ba
&&\Delta z(\Omega_B h^2,\Omega_0h^2)= 0.5285 \,\left[(\Omega_B h^2)^{-1}
-0.0125^{-1}\right] \left(\frac{\Omega_0 h^2}{0.25}\right)^{0.31}
\nonumber\\ && \hskip 4cm
 -7.022\times 10^{-4} \, \left[(\Omega_B h^2)^{-1}-0.0125^{-1}\right]^2
\left(\frac{\Omega_0 h^2}{0.25}\right)^{0.55}
\nonumber\\
&&\Delta z(\Omega_0 h^2)= 73.21 \left[\sqrt{\Omega_0 h^2}-0.5\right]
 -12.06\, \left[\sqrt{\Omega_0 h^2}-0.5\right]^2 \nonumber\\
&&\Delta z(N_{\nu},\Omega_B h^2, \Omega_0 h^2)= 0.3823 (N_{\nu}-3)
\left(\frac{\Omega_B h^2}{0.0125}\right)^{-0.756}
\left(\frac{\Omega_0 h^2}{0.25}\right)^{-0.46}.
\ea
Our fitting formula is accurate to $\Delta z=\pm 1$ for the parameter
ranges of $ 0.1 \leq \Omega_0 h^2 \le 0.64$, $0.005 \leq \Omega_B h^2
 \leq 0.03$, and $2\leq N_{\nu} \leq 12$.

\newpage
\nonfrenchspacing
\parindent=20pt
\centerline{{\bf Figure Captions}}

Fig.1 The predicted scalar and tensor contributions. The set
of cosmological parameters is: $h=0.5$, $\Omega_B=0.05$, $N_\nu=3$,
$\Omega_\Lambda=0$, $n_T=0$.

Fig.2 Histograms of maximum error for $l\leq 50$ and $l\leq 100$
($\Omega_\Lambda =0$).

Fig.3 Histograms of maximum error for $l\leq 50$ and $l\leq 100$
($\Omega_\Lambda >0$).

\newpage
\begin{figure}
\psfig{file=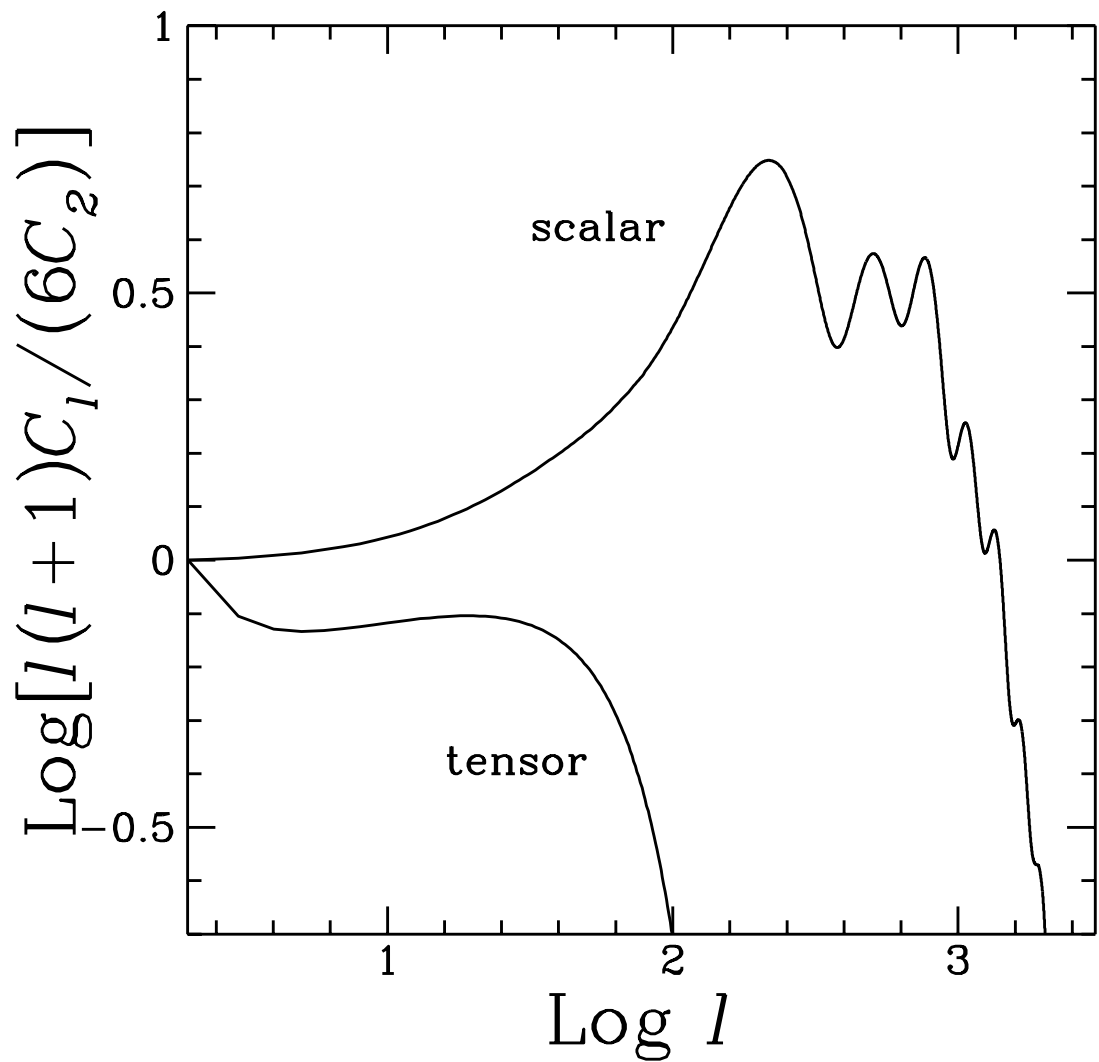,height=6in,width=6in}
\caption{The predicted scalar and tensor contributions. The set
of cosmological parameters is: $h=0.5$, $\Omega_B=0.05$, $N_\nu=3$,
$\Omega_\Lambda=0$, $n_T=0$.}
\end{figure}

\begin{figure}
\psfig{file=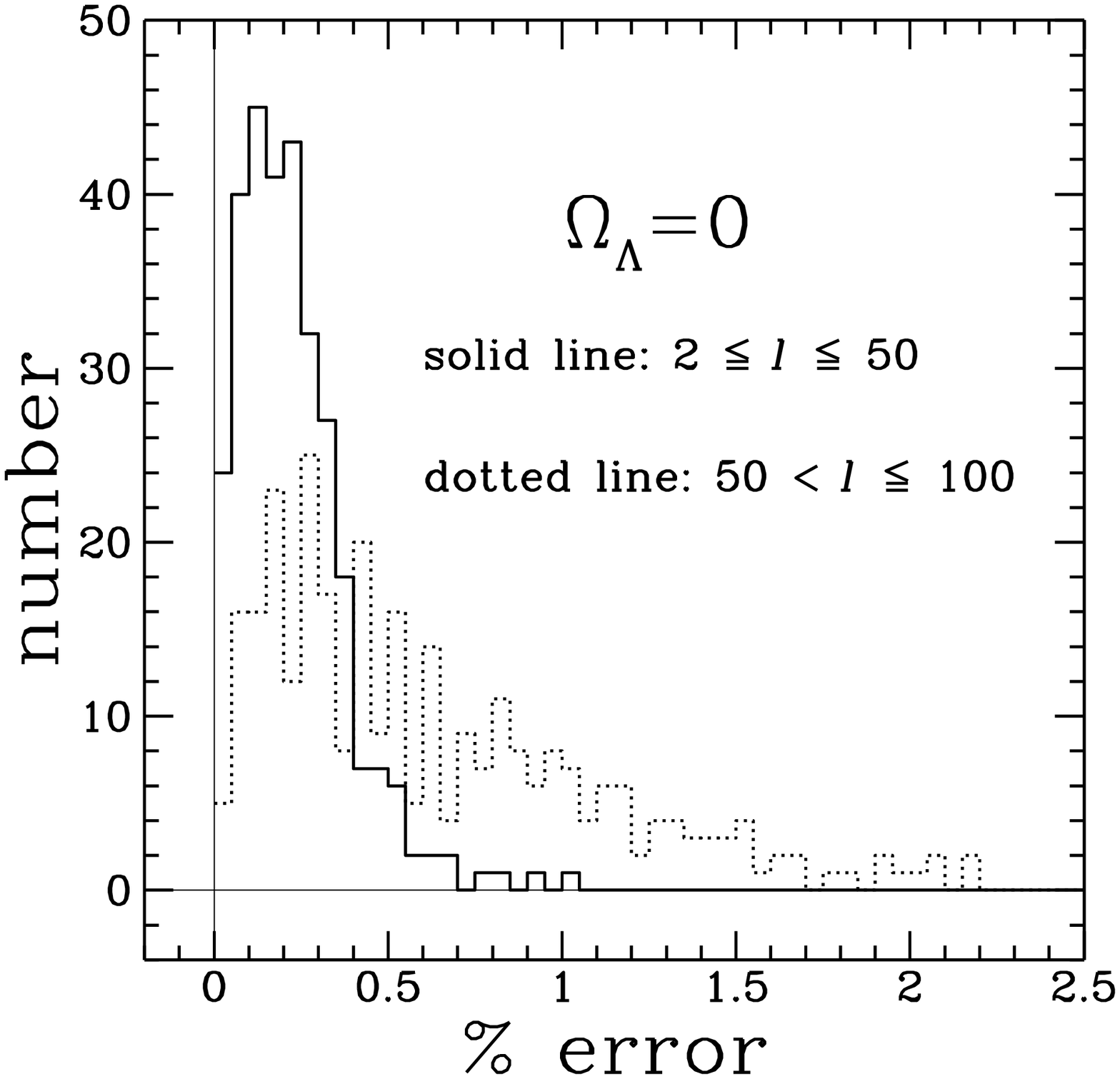,height=6in,width=6in}
\caption{Histograms of maximum error for $l\leq 50$ and $l\leq 100$
($\Omega_\Lambda =0$).}
\end{figure}

\begin{figure}
\psfig{file=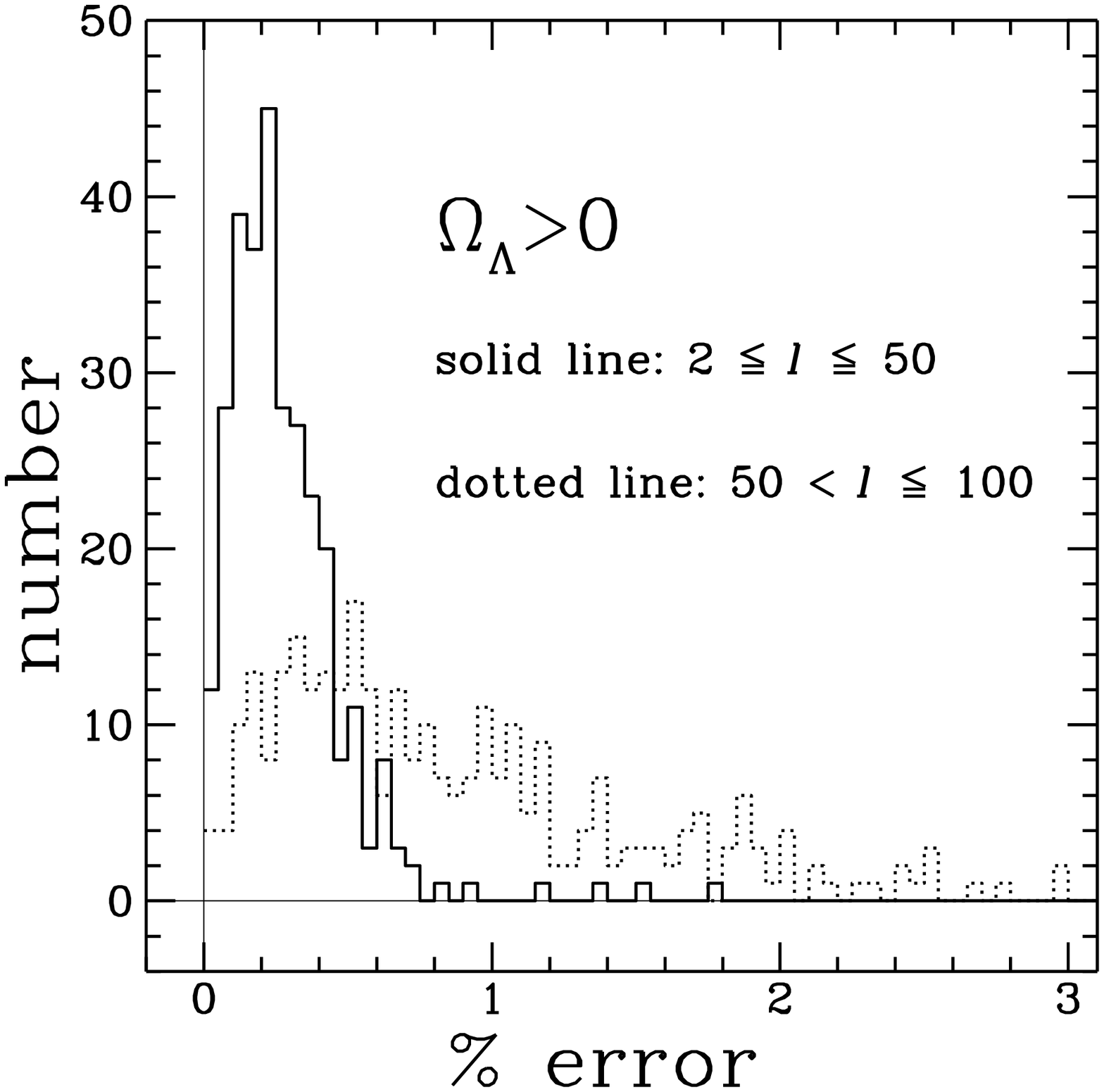,height=6in,width=6in}
\caption{Histograms of maximum error for $l\leq 50$ and $l\leq 100$
($\Omega_\Lambda >0$).}
\end{figure}



\end{document}